\documentclass[epj]{webofc}
\usepackage[utf8]{inputenc}
\usepackage[varg]{txfonts}   
\usepackage{booktabs}
\usepackage{xcolor}
\definecolor{darkred}{rgb}{0.4,0.0,0.0}
\definecolor{darkgreen}{rgb}{0.0,0.4,0.0}
\definecolor{darkblue}{rgb}{0.0,0.0,0.4}
\usepackage[bookmarks,linktocpage,colorlinks,
    linkcolor = darkred,
    urlcolor  = darkblue,
    citecolor = darkgreen]{hyperref}
 
\usepackage{physics}
\usepackage{tabularx} 
\usepackage{davitex}  
 
\wocname{EPJ Web of Conferences}
\woctitle{Lattice2017}

\begin{document}
\selectlanguage{english}

\title{Estimates of Scaling Violations for Pure SU(2) LGT}

\author{%
\firstname{Bernd} \lastname{Berg}\inst{1} \and
\firstname{David} \lastname{Clarke}\inst{1}\fnsep\thanks{Speaker,
\email{dclarke@hep.fsu.edu}}
}

\institute{Florida State University Department of Physics, 
           Tallahassee, FL 32306}

\abstract{%
We investigate the approach of pure SU(2) lattice gauge theory with the 
Wilson action to its continuum limit using the deconfining transition, 
L\"uscher's gradient flow~\cite{luscher_properties_2010}, and the 
cooling flow \cite{berg_dislocations_1981,bonati_comparison_2014} to set 
the scale. Of those, the cooling flow turns out to be computationally 
most efficient. We explore systematic errors due to use of three 
different energy observables and two distinct reference values for 
the flow time, the latter obtained by matching initial scaling behavior 
of some energy observables to that of the deconfining transition. Another 
important source of systematic errors are distinct fitting forms for the 
approach to the continuum limit. Besides relying in the conventional way 
on ratios of masses, we elaborate on a form introduced by Allton 
\cite{allton_lattice_1997}, which incorporates 
asymptotic scaling behavior. Ultimately we find that, though still 
small, our systematic errors are considerably larger than our 
statistical errors.}

\maketitle

\section{Introduction}\label{sec:intro}

Tests of the Standard Model, and New Physics searches require 
precise predictions of physical observables and accurate estimates 
of systematic uncertainty. Since scale setting is a source of 
uncertainty, it is useful to investigate scales that can achieve 
small statistical error bars, and to estimate the systematic error 
picked up from the choice of scale and the choice of fitting form 
for continuum limit extrapolations. Pure SU(2) gauge theory is a 
good testing ground for new scales because one can generate large 
statistics with relatively modest CPU resources. 

Here we summarize the results \cite{berg_clarke_2017} of a study of 
pure SU(2) LGT with the Wilson action
\begin{equation}
  S=\beta\sum_\Box\left(1-\frac{1}{2}\Tr
      U_\Box\right),~~~~\beta=\frac{4}{g^2},
\end{equation}
where $g$ is the bare coupling constant and the summation runs over 
all plaquettes. We parameterize lattice expectation values of plaquette 
matrices by
\begin{equation} \label{eq:par}
  \ev{U_\Box}_L=a_0\id+i\sum_{i=1}^3a_i\sigma_i\,.
\end{equation}
With this parameterization we use three definitions of the energy 
density: 
\begin{equation} \label{eq:energy}
  E_0\equiv 2\,(1-a_0),~~~~
  E_1\equiv\sum_{i=1}^3 a_i^2,~~~~\text{and}~~~~
  E_4\equiv\frac{1}{16}\sum_{i=1}^3
           \left(a_i^{ul}+a_i^{ur}+a_i^{dl}+a_i^{dr}\right)^2,
\end{equation}
where $E_4$ is L\"uscher's energy density, which averages over 
the four plaquettes attached to each site $x$ in a fixed $\mu\nu$, 
$\mu\ne\nu$ plane; the superscripts of $a_i$ stand 
for up ($u$), left ($l$), right ($r$), and down ($d$) with 
respect to $x$ (drawn in figure~1 of \cite{luscher_properties_2010}).
All definitions become $\sim F_{\alpha\beta}F_{\alpha\beta}$ 
in the continuum limit.

Configurations are obtained by Markov chain Monte Carlo
(MCMC) simulations with Monte Carlo plus Overrelaxation (MCOR) updating.
One MCOR sweep updates each link once in a systematic order 
with the Fabricius-Haan-Kennedy-Pendleton 
\cite{fabricius_heat_1984,kennedy_improved_1985} heatbath algorithm and, 
in the same systematic order, twice by overrelaxation 
\cite{adler_overrelaxation_1988}. Using checkerboard coding 
\cite{barkai_can_1982} and MPI Fortran, parallel updating of 
sublattices is implemented.

\section{Reference scales}\label{sec:scales}

\subsection{Deconfinement scale}
The first scale investigated is the deconfinement scale. It is used
to guide our choice of target values for the other scales. To calculate 
this scale we perform simulations on $N_s^3N_{\tau}$ lattices and 
estimate pseudocritical coupling constants $\beta_c(N_s,N_\tau)$ by 
reweighting the Polyakov loop susceptibility
\begin{equation} \label{eq:chi}
  \chi(\beta)=\frac{1}{N_s^3}\left(\ev{P^2}-\ev{|P|}^2\right),~~~~
   P=\sum_{\vec{x}}\prod_{x_4}U(\vec{x},x_4)
\end{equation}
to nearby $\beta$ values and finding the maximum. Critical coupling 
constants $\beta_c(N_{\tau})=\beta_c(\infty,N_\tau)$ 
are then extrapolated using three-parameter fits 
\begin{equation} \label{eq:bc}
  \beta_c(N_s,N_{\tau})=\beta_c(N_{\tau}) 
  +a_1(N_{\tau})N_s^{a_2(N_{\tau})}.
\end{equation}
Inverting the results of these fits defines 
the deconfining length scale $N_{\tau}(\beta_c)$.
Configurations were generated using $2^{19}-2^{25}$ MCOR sweeps,
which were partitioned into 32 or more bins. Error bars are then
calculated using the jackknife procedure on these bins.

\begin{figure}[b]
  \centering
  \includegraphics[width=0.489\linewidth]{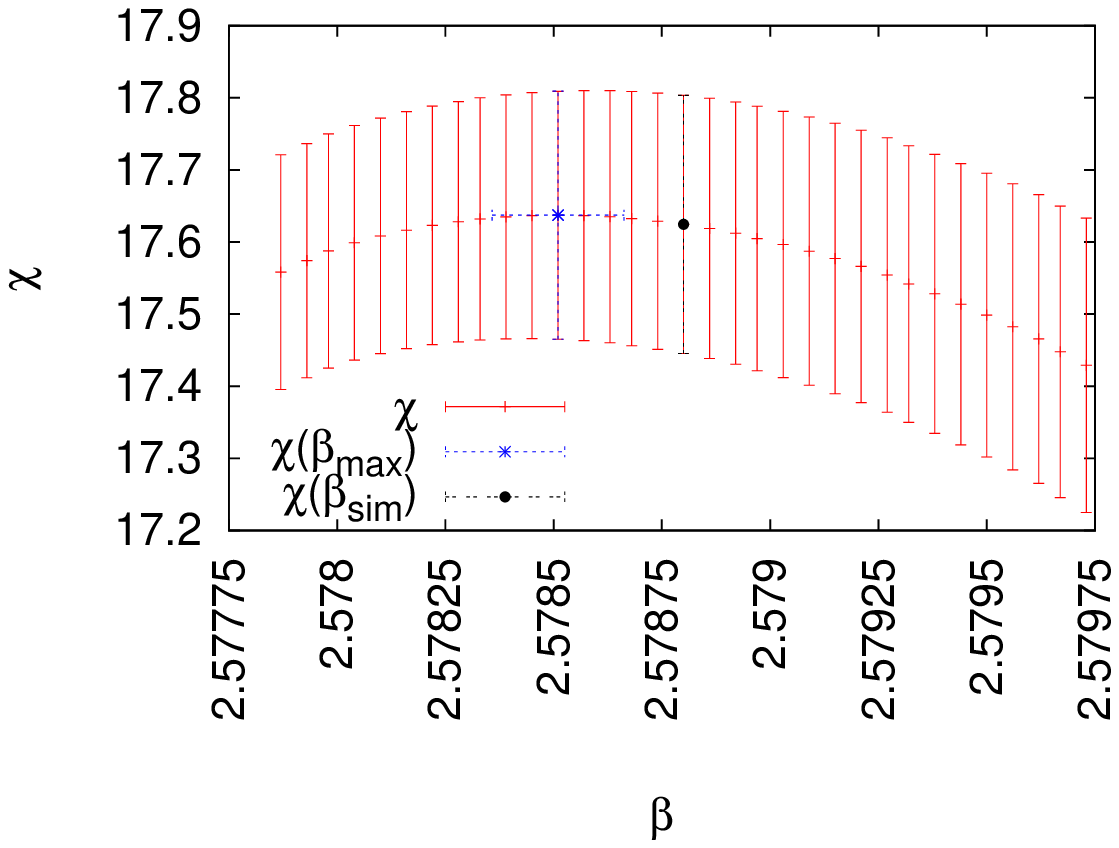}
  \includegraphics[width=0.489\linewidth]{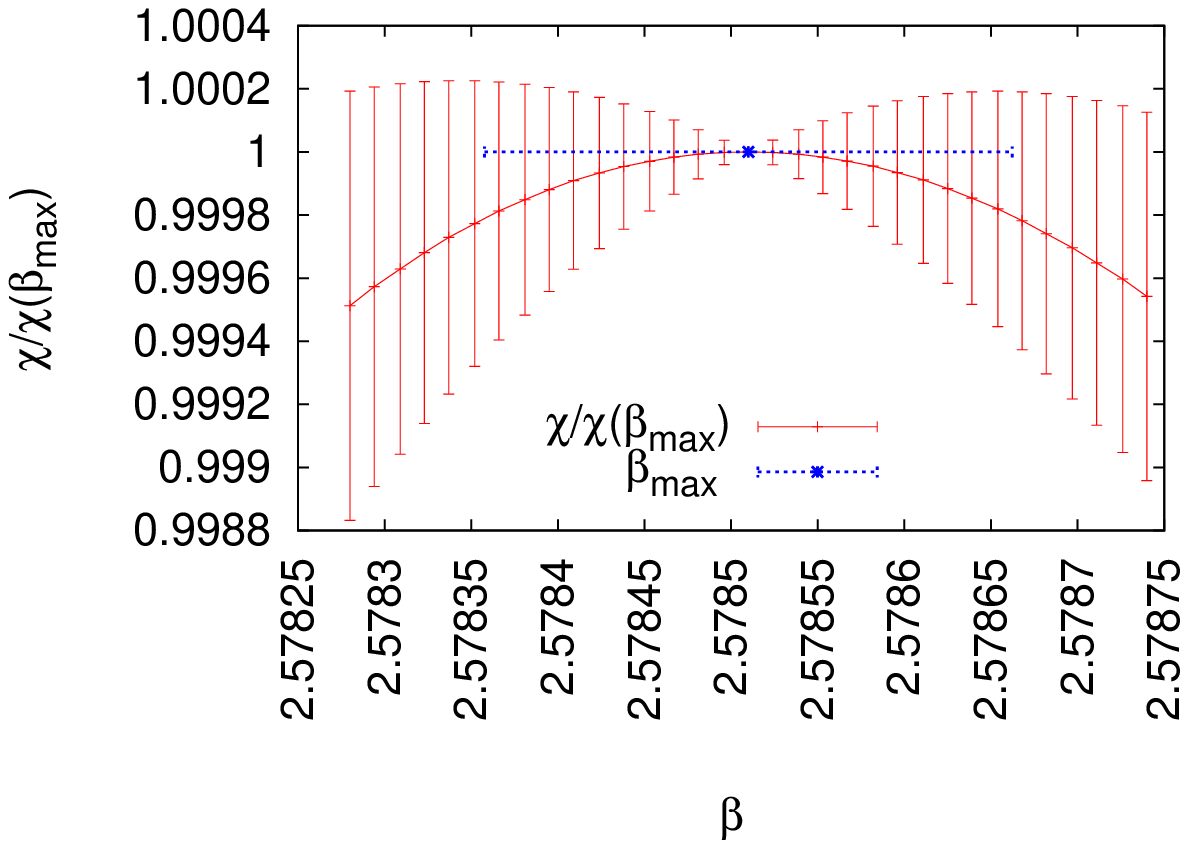}
  \caption{Left: Reweighted susceptibility curve on a $64^3\times 10$
           lattice simulated at $\beta=2.5788$. Right:
           Susceptibility curve with maximum value divided out.}
  \label{fig:chirat}
\end{figure}

The reweighting curve for the susceptibility is rather flat near
the maximum and has large error bars, as shown in figure~\ref{fig:chirat} 
(left). Nevertheless the error bar on the position 
of the pseudocritical beta is relatively small. This is explained 
by the fact that the error bars of figure~\ref{fig:chirat} (left) 
are strongly correlated, since they are reweighted from the same 
simulation. Dividing out the maximum susceptibility in each jackknife 
bin gives us figure~\ref{fig:chirat} (right), which makes
the rather small error bar of $\beta_{\max}$ plausible.

\subsection{Gradient scale}
The gradient flow equation \cite{luscher_properties_2010}
\begin{equation}\label{eq:gflow}
  \dot{V}_\mu(x,t)=-g^2V_\mu(x,t)\partial_{x,\,\mu}S[V(t)]
\end{equation}
is an evolution equation in the fictitious flow time $t$ that
decreases the action as $t$ increases.
The SU(2) link derivatives are defined by
\begin{equation}
  \partial_{x,\,\mu}f(V)\equiv 
     i\sum_{i=1}^3\sigma_i\frac{d}{ds}f(e^{isX^i}V)\big|_{s=0},~~~~
     X^i(x',\mu')=
     \begin{cases}
       ~\sigma^i & \text{ if $(x',\mu')=(x,\mu)$},\\
       ~0       & \text{ otherwise},
     \end{cases}
\end{equation}
and the initial condition is $U_\mu(x,0)=U_\mu(x)$.
The functions
\begin{equation} \label{eq:y}
  y_i(t)=t^2E_i(t),~~~~i=0,1,4
\end{equation}
are used to define gradient scales by choosing appropriate fixed 
{\it target values} $y_i$ and integrating the gradient flow equation
until eq.~\eqref{eq:y} is satisfied. As a function of $\beta$, 
the observable
\begin{equation} \label{eq:s}
  s_i(\beta)=\sqrt{t_i(\beta)}
\end{equation}
scales like a length provided that
\begin{enumerate}
  \item lattice sizes are chosen so that $N_{\min}\gg \sqrt{8}s_i$, 
        where $\sqrt{8}\,s_i$ is the smoothing range 
        \cite{luscher_properties_2010} and $N_{\min}=\min\{N_i:i=1,2,3,4\}$ 
        for simulations on a $N_1N_2N_3N_4$ lattice;
  \item the values of $\beta$ are large enough to be in the 
        SU(2) scaling region; and
  \item the target values are large enough so that $\sqrt{8}
        s_i\gg 1$ for the smallest used flow time.
\end{enumerate}

\begin{figure}[b]
  \centering
  \includegraphics[width=0.489\linewidth]{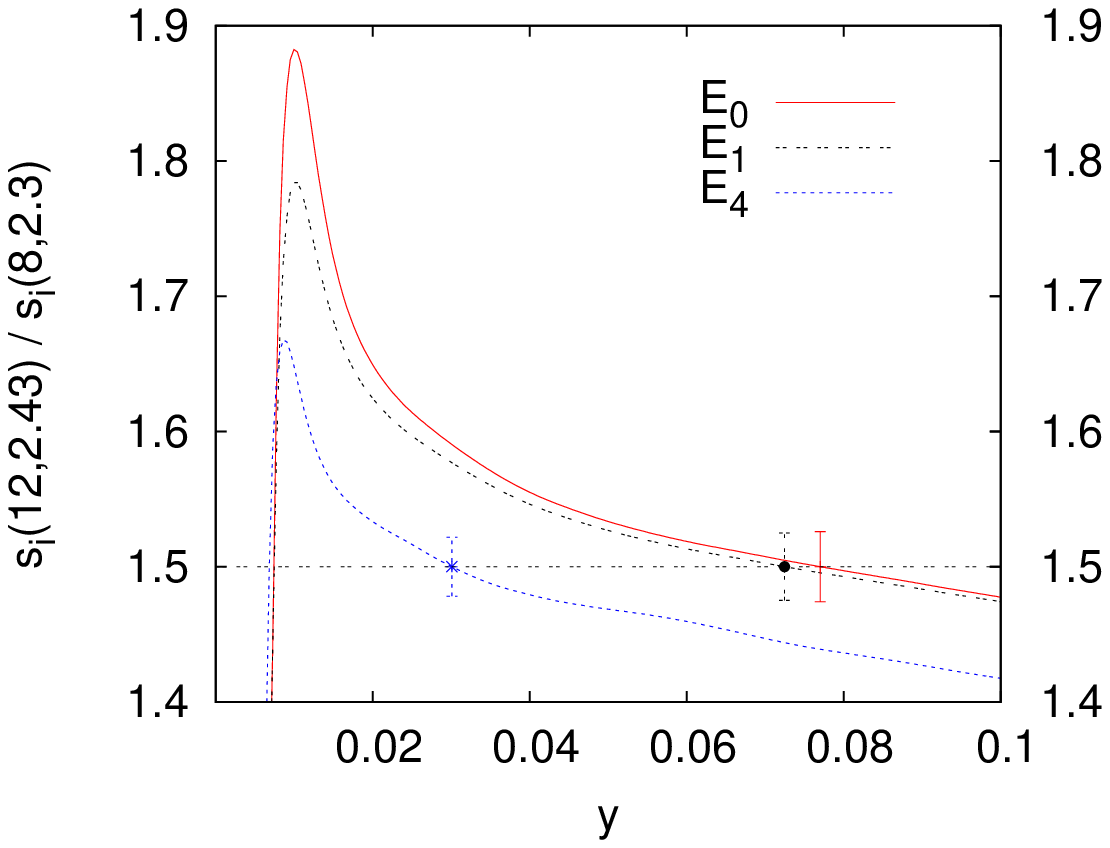}
  \includegraphics[width=0.489\linewidth]{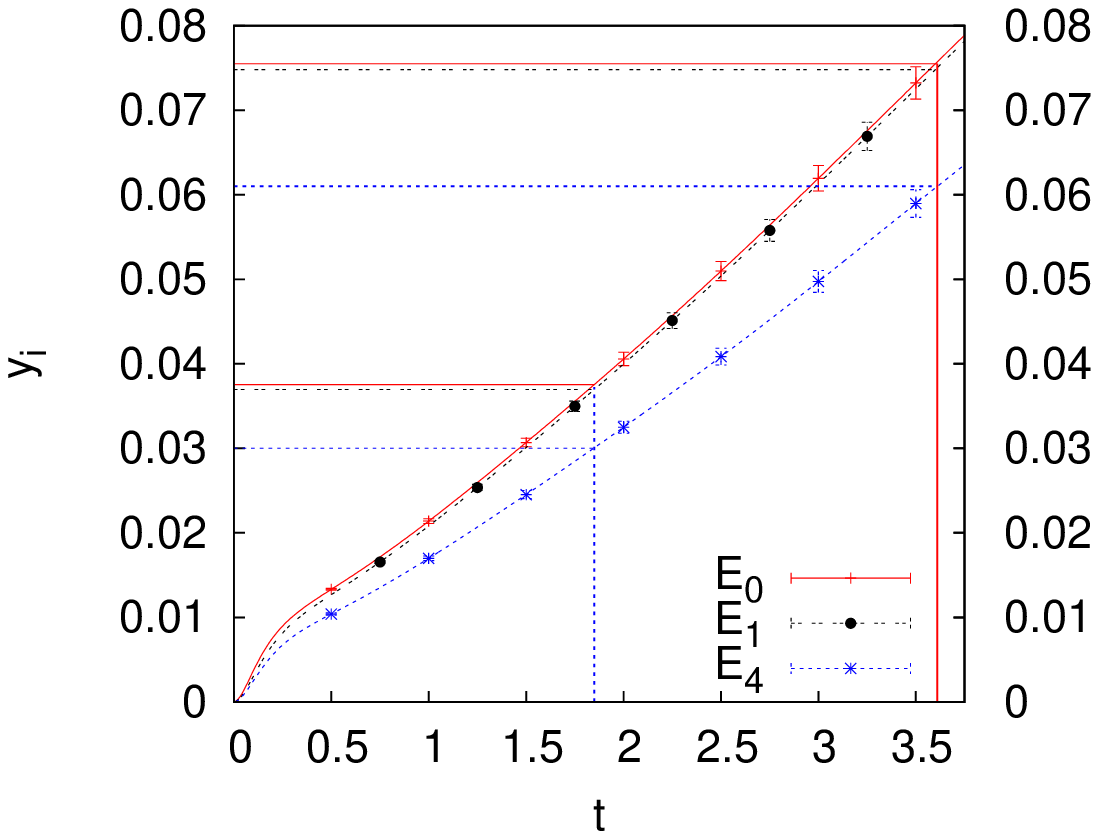}
  \caption{Left: Gradient flow ratios as function of $y$.
           Right: Gradient flow of an $8^4$ lattice at $\beta=2.3$.}
  \label{fig:gfrat}
\end{figure}

One can choose target values so that initial estimates of the scales 
$s_i$ agree with the deconfinement scale for low values of $\beta$. 
For example figure~\ref{fig:gfrat} (left) plots the ratio of 
$s_i\,(N=12,\,\beta=2.43)$ to $s_i\,(N=8,\,\beta=2.30)$ against 
$y_i$. Due to scaling, one expects this ratio to agree with the 
ratio of the deconfinement scales, $N_\tau(2.43)/N_\tau(2.3)=12/8=1.5$. 
This leads to two intersections, one coming from the $E_4$ curve and 
another coming from the $E_0$ and $E_1$ curves, which practically agree.
Figure~\ref{fig:gfrat} (right) plots the function $t^2 E_i$ against 
the flow time. Picking initially $y_4^1=0.030$, the target value 
corresponding to the aforementioned $E_4$ intersection, defines a 
flow time, indicated by the vertical dotted blue line at $t=1.85$. 
This flow time is then used to define two more target values $y_0^1$ 
and $y_1^1$ using that figure. Similarly, picking initially $y_0^2=
0.0755$ (or equivalently $y_2^1=0.0748$) delivers three more targets. 
Hence, we consider altogether six target values: 
\begin{gather} \label{eq:gfy1}
   y^1_0=0.0376,~~y^1_1=0.0370,~~y^1_4=0.030,\\
   \label{eq:gfy2}
   y^2_0=0.0755,~~y^2_1=0.0748,~~y^2_4=0.061.
\end{gather}

\subsection{Cooling scale}\label{sec:coolingscale}
The cooling algorithm was introduced as part of an investigation of 
topological charge in the 2D O(3) sigma model \cite{berg_dislocations_1981}. 
Since then it has found many applications \cite{vicari_panagopoulos_2009}. 
Bonati and D'Elia showed that using cooling as a smoothing technique produces
similar results for topological observables as the gradient flow
\cite{bonati_comparison_2014}. 
An SU(2) cooling step updates a link variable by
\begin{equation}\label{eq:cool}
  V_\mu(x,n_c)=\frac{V^\sqcup_\mu(x,n_c-1)}{|V_\mu^\sqcup(x,n_c-1)|},
\end{equation}
where $n_c$ is the number of cooling steps and $V_\mu^\sqcup(x)$ 
is the staple matrix
\begin{equation}
  V_\mu^\sqcup(x) =\sum\limits_{\nu\neq\mu}\left[
                    V_\nu(x)V_\mu(x+\hat{\nu})V_\nu^\dagger(x+\hat{\mu})
                  + V_\nu^\dagger(x-\hat{\nu})V_\mu(x-\hat{\nu})
                    V_\mu(x-\hat{\nu}+\hat{\mu})\right].
\end{equation}
The update \eqref{eq:cool} minimizes the local contribution to the 
action, so that the "cooling flow" decreases the action. A cooling 
scale is then defined in the same way as a gradient scale through 
eq.~\eqref{eq:s} by iterating eq.~\eqref{eq:cool} until a specified 
target value is reached. In 4D, $n_c$ cooling steps correspond to a 
gradient flow time $t_c=n_c/3$ \cite{bonati_comparison_2014}; therefore 
the cooling flow attains its target value at least 34 times faster than 
the gradient flow, assuming eq.~\eqref{eq:gflow} is integrated using 
the Runge-Kutta scheme with $\epsilon=0.01$. Matching cooling scale 
ratios to deconfinement scale ratios, the target values are 
\begin{gather} \label{cyi01}
  y^1_0=0.0440,~~y^1_1 = 0.0430,~~y^1_4 = 0.0350,\\
  y^2_0=0.0822,~~y^2_1 = 0.0812,~~y^2_4 = 0.0656.
\end{gather}

\section{Data generation for gradient and cooling scales}\label{sec:data}
Lattice sizes and $\beta$ values used are given in 
Table~\ref{tab:sizes}. In each run, 128 configurations were generated, 
and on each configuration the gradient and cooling flows were performed. 
We allocated approximately equal amounts of CPU time to configuration 
generation and to smoothing flow. Subsequent configurations are separated 
by $2^{11}-2^{13}$ MCOR sweeps, depending on how many sweeps are needed 
to reach the target value. Error bars are calculated using the jackknife
method with respect to the 128 configurations. 

\begin{table}[thb]
  \centering
  \caption{Lattice size used to generate data at each $\beta$ value.}
  \label{tab:sizes}
  \begin{tabular}{ll}\toprule
  Lattice Size  & $\beta$ values  \\\midrule
  $16^4$ & 2.300\\
  $28^4$ & 2.430, 2.510\\
  $40^4$ & 2.574, 2.620, 2.670, 2.710, 2.751\\
  $44^4$ & 2.816\\
  $52^4$ & 2.875\\\bottomrule
  \end{tabular}
\end{table}

\begin{figure}[b]
  \centering
  \includegraphics[width=8.7cm]{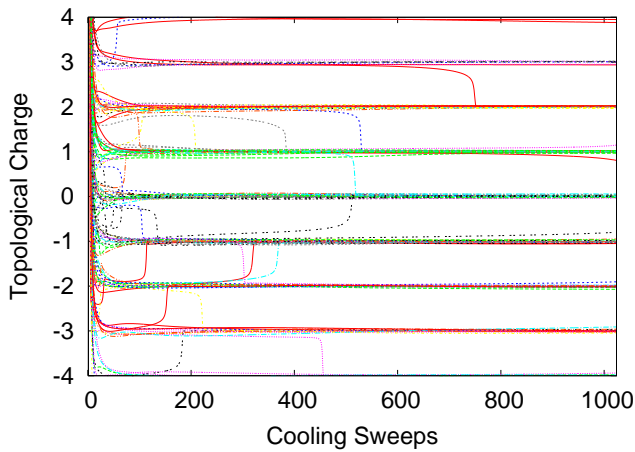}
  \caption{Cooling of the topological charge on a $44^4$ lattice
           at $\beta=2.816$.}
  \label{fig:topcool}
\end{figure}

Integrated autocorrelation times $\tau_\text{int}$ for the series 
of configurations are estimated using the software of reference 
\cite{berg_markov_2004} and are found to be statistically compatible 
with~1. Additionally we calculated for each configuration the cooling 
flow of the topological charge defined as in reference 
\cite{bonati_comparison_2014}. An example of cooling trajectories 
for the 128 configurations is given in figure~\ref{fig:topcool}. 
Topological correlations are then estimated by calculating 
$\tau_\text{int}$ for the series of topological charges, which
are also found to be statistically compatible with 1. Thus we
treat our scale measurements as statistically independent.

\section{Scaling and asymptotic scaling analysis}\label{sec:scaling}
We analyze 13 length scales, the deconfinement scale $L_0$, the gradient
scales $L_1-L_6$ and the cooling scales $L_7-L_{12}$. Our goals are to 
determine whether using cooling scales over gradient scales leads to a 
significant loss of accuracy; estimate scaling violations; investigate 
whether the choice of target value leads to seriously distinct scaling 
behavior; compare scaling and asymptotic scaling fit forms; and finally 
give an overall estimate of systematic error picked up from reference 
scale choice and choice of fitting form for continuum limit 
extrapolations.

\begin{figure}[t] 
  \centering
  \includegraphics[width=0.489\linewidth]{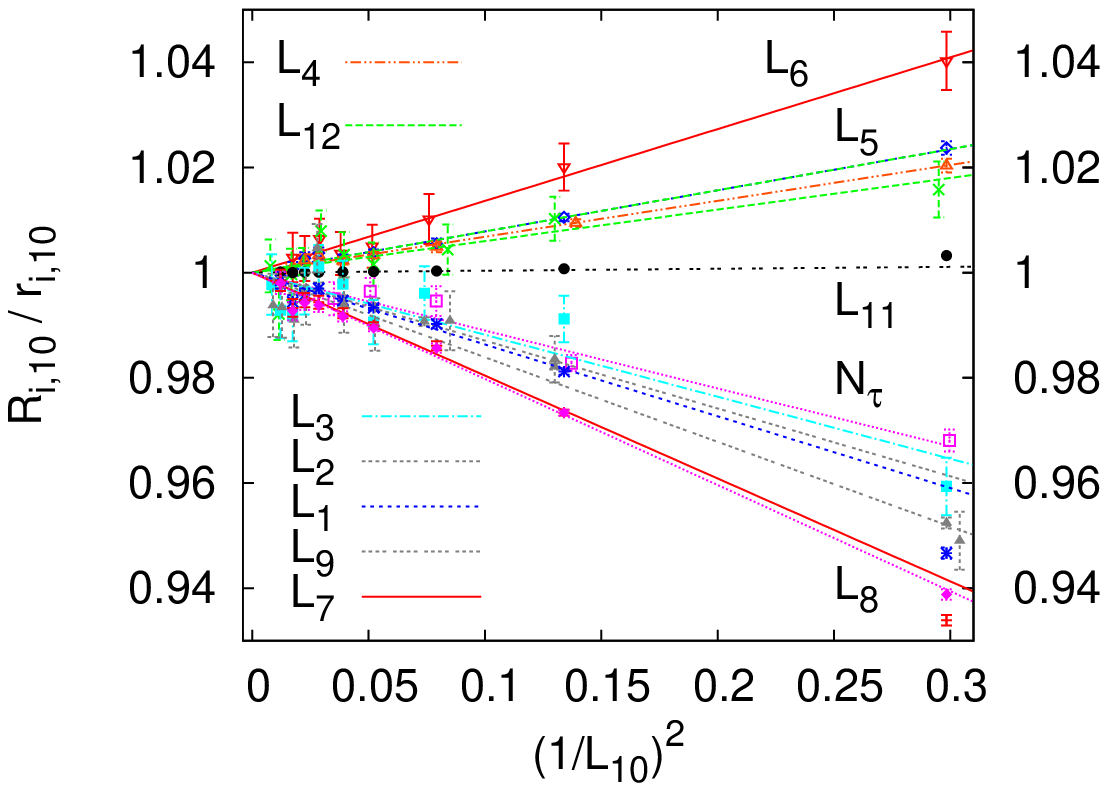}
  \includegraphics[width=0.489\linewidth]{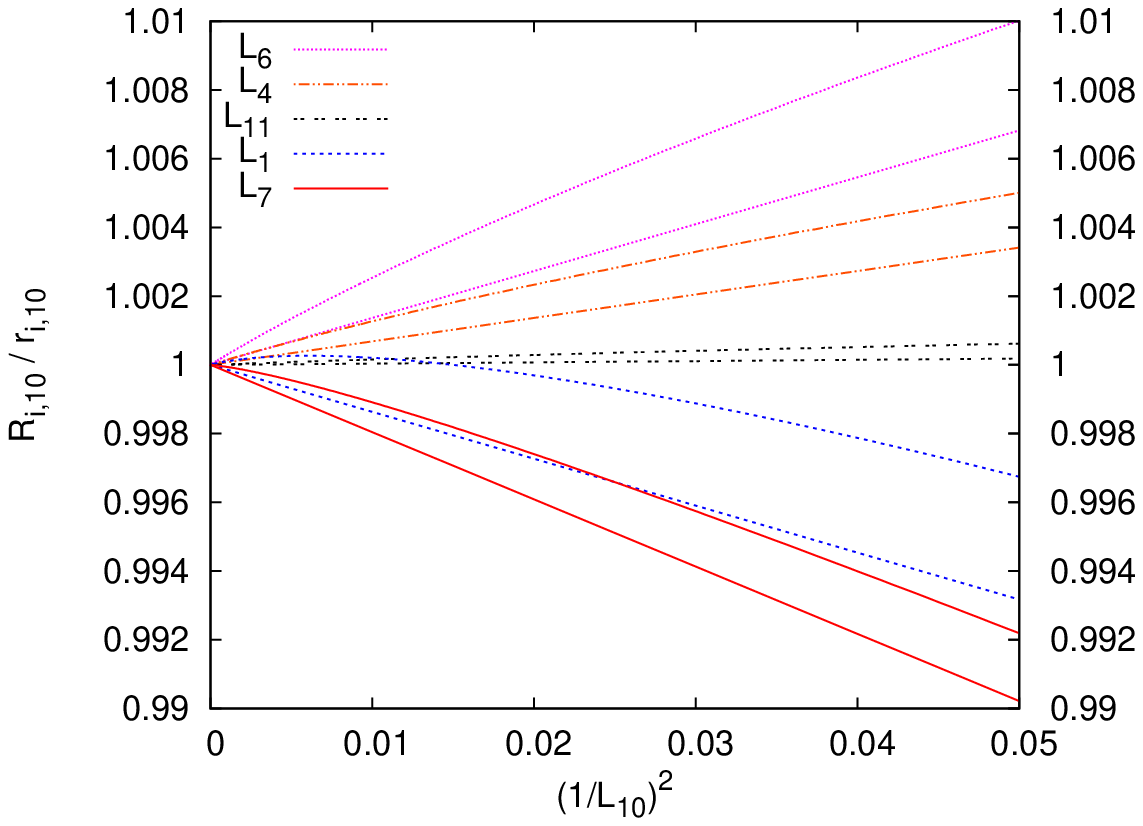}
  \caption{Left: Scaling corrections for ratios $L_i/L_{10}$. Some 
           labels are attached to lines and others are in the legend. 
           Here and in the remaining plots the top-bottom ordering of 
           the legend matches the top-bottom ordering of the plot, 
           and some data are shifted slightly for visibility. Right:
           Direct comparison between representative scaling fits from
           the left plot and asymptotic scaling fits in the same region.
           Slightly curved fits of the pairs belong to the asymptotic 
           scaling form, and the data are omitted for visibility.}
  \label{fig:LiLj}
\end{figure}

We first examine 
$\mathcal{O}(a^2)$ scaling corrections to ratios of lengths. These
are fitted using
\begin{equation}
  R_{i,j}\equiv\frac{L_i}{L_j}\approx r_{i,j}+k_{i,j}a^2\Lambda_L^2
         =r_{i,j}+c_{i,j}\left(\frac{1}{L_j}\right)^2,
\end{equation}
where $r_{i,j}$ is the continuum limit value and $k_{i,j}$ and $c_{i,j}$
are constants. In the following we fix $L_j=L_{10}$ and plot the
normalized ratio
\begin{equation}
  \frac{R_{i,10}}{r_{i,10}}=1+c'_{i,10}\left(\frac{1}{L_{10}}\right)^2
\end{equation}
in figure~\ref{fig:LiLj} (left). We find no discernible loss 
of accuracy using the cooling scale over the gradient scale, which one 
can see from relative sizes of error bars in this plot. 
There is clear overlap between cooling and gradient scales, for
example cooling scales $L_{10}-L_{12}$ fall within the spread of
gradient scales $L_1-L_6$, which shows that cooling scales do not 
suffer significant scaling violations compared to gradient scales. 
At $(1/L_{10})^2\approx0.30$, which corresponds to $\beta=2.300$, 
we read off scaling violations of about 10\%. Deeper in the scaling 
region at $(1/L_{10})^2\approx0.05$, which corresponds to 
$\beta=2.574$, the violations are less than 2\%. Results for all 
our $L_i/L_j$ continuum estimates are compiled in Table~\ref{tab}.

\begin{table}[t]
  \centering
  \caption{Estimates of the continuum limits of the 
           $L_i/L_j$ ratios.} 
  \label{tab}
  \begin{tabular}{lllll}\toprule
  $i\,\backslash\, j$& $L_1$ & $L_4$  & $L_7$ & $L_{10}$ \\ \midrule
  $L_0$   & 2.8896(71) & 2.2290(46) & 2.8855(68) & 2.2618(42) \\
  $L_1$   &            & 0.77382(61)& 0.99845(38)& 0.78433(43)\\
  $L_3$   & 0.9250(19) & 0.7163(17) & 0.9241(19) & 0.7264(16) \\
  $L_4$   & 1.2943(11) &            & 1.29135(99)& 1.01520(49)\\
  $L_6$   & 1.2090(26) & 0.9346(20) & 1.2081(27) & 0.9490(21) \\
  $L_7$   & 1.00156(38)& 0.77398(79)&            & 0.78570(50)\\
  $L_9$   & 0.9222(21) & 0.7141(19) & 0.9213(20) & 0.7243(17) \\
  $L_{10}$& 1.27509(70)& 0.98508(47)& 1.27300(80)&            \\
  $L_{12}$& 1.1835(24) & 0.9164(21) & 1.1825(24) & 0.9292(19) \\
  \bottomrule
  \end{tabular}
\end{table}

\begin{figure}[t]
  \centering
  \includegraphics[width=10cm]{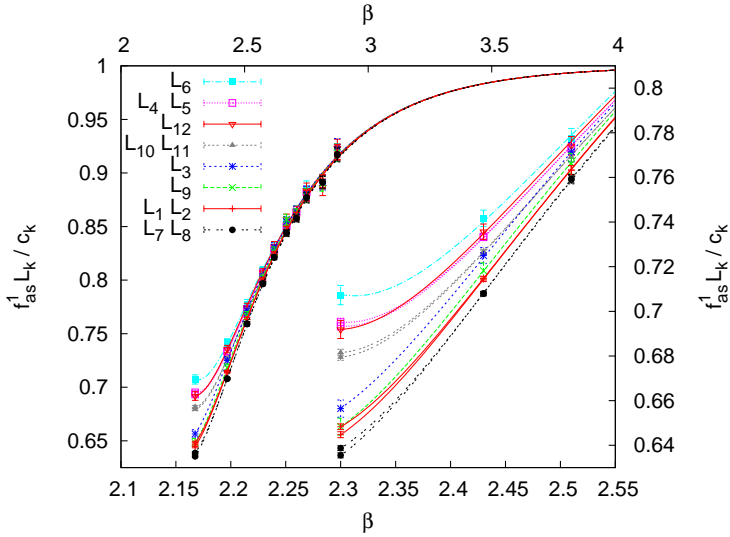}
  \caption{Asymptotic scaling corrections for lengths $L_k$. The top 
     abscissa and left ordinate correspond to 
     the top set of curves, while the bottom abscissa and right ordinate 
     correspond to the bottom set of curves, which is an enlargement of 
     the top curves for low $\beta$ values.}
  \label{fig:ascaling}
\end{figure}

The three-loop asymptotic scaling relation for SU(2) is
\begin{equation}\label{eq:ascale}
  a\Lambda_L\approx\exp\left(-\frac{1}{2b_0g^2}\right)
                  (b_0 g^2)^{-b_1/2b_0^2}
                  \left(1+ q_1g^2\right)\equiv f_{as}^1(\beta),
\end{equation}
where $b_0=11/24\pi^2$ \cite{gross_asymptotically_1973,politzer_reliable_1973}
and $b_1=17/96\pi^4$ \cite{caswell_asymptotic_1974,jones_two-loop_1974} 
are the regularization scheme independent coefficients and $q_1=0.08324$ 
\cite{alle_three-loop_1997} using lattice regularization. The superscript 
of $f_{as}^1$ indicates that only the $\mathcal{O}(g^2)$ correction to 
the asymptotic scaling relation is included. Allton suggested 
\cite{allton_lattice_1997} using eq.~\eqref{eq:ascale} to fit the approach 
to the continuum limit by expanding in powers of $a$. For the inverse 
length $1/L_i$ this reads
\begin{equation}\label{eq:expand}
      \frac{1}{L_i}=c'_ia\Lambda_L
          \left(1+\sum_{j=1}^\infty\alpha'_{i,j}(a\Lambda_L)^j\right).
\end{equation}
Plugging eq.~\eqref{eq:ascale} into eq.~\eqref{eq:expand} and truncating
the series after 3 terms, which is essentially the minimum number needed 
to get acceptable $q$ values for asymptotic scaling fits, one finds
\begin{equation}\label{eq:ascaling}
    L_i\approx\frac{c_i}{f^1_{as}(\beta)}
      \left(1+\sum_{j=1}^3\alpha_{i,j}[f_{as}^1(\beta)]^j\right).
\end{equation}
Due to the $\alpha_{i,1}$ term in eq.~\eqref{eq:ascaling} 
corrections to ratios would in general be of order $a$ in the lattice
spacing. This may be the main reason why Allton's approach never became
popular. In \cite{berg_asymptotic_2015} this problem was avoided by
combining the scales discussed there into a single fit, which is only
possible if the relative scaling violations are so weak that they can be
neglected within the statistical errors. In \cite{berg_clarke_2017}
we relaxed this to the requirement that the $\alpha_{i,1}$ coefficients
have to agree for all scales, i.e. $\alpha_{i,1}\equiv \alpha_1$, 
which provides a general solution to the problem. The master 
coefficient was then determined by the maximum likelihood approach, 
varying $\alpha_1$ and repeating all fits for each value.
Figure~\ref{fig:ascaling} plots eq.~\eqref{eq:ascaling} against
$\beta$, with the asymptotic scaling behavior divided out. With 
this normalization the curves approach 1 in the continuum limit. 
At $\beta=2.300$ asymptotic scaling violations range from 28\% 
to 37\%. Differences of ratios at this $\beta$ value reach 14\%
(from $0.72/0.63\approx1.14$), in agreement with 12\% (from 
$1.04/0.93\approx 1.12$) of the scaling fits at the same $\beta$ 
value, shown in figure~\ref{fig:LiLj} (left).

For a more direct comparison with scaling, we compute ratios of gradient 
and cooling lengths using asymptotic scaling. Since all scales have the
same first order coefficient, they cancel in the ratio, leading to 
normalized ratio functions
\begin{equation}
  \frac{R_{i,10}}{r_{i,10}}=1+\frac{1}{r_{i,10}}
     \sum_{j=2}^3\kappa_{i,j}\left[f^1_{as}\right]^j.
\end{equation}
In the figure~\ref{fig:LiLj} (right), normalized ratio functions 
using scaling and asymptotic scaling fits are both plotted against the 
squared lattice spacing. Straight line fits are scaling fits, while 
slightly curved fits are asymptotic scaling fits. Systematic errors 
due to choice of fitting form alone seems not to exceed about 0.6\%.
At $(1/L_{10})^2=0.05$, the combined systematic error due to choice 
of scale, target value, and continuum limit fitting form is read off 
to be around 2\%.

\section{Summary and Conclusions}\label{sec:summary}
A physical scale, here the deconfinement scale, is well suited for
determining initial target values for the cooling and gradient flow.
The cooling scale is found to reach its 
target values much faster than the gradient scale without losing 
accuracy in scale setting. Systematic errors due to using distinct
operators, within the same scale, gradient or cooling, are considerably
larger than the systematic error encountered by switching from the
gradient to the cooling scale or vice versa. For the gradient and 
cooling scales six target values were used, and the choice of target value
gives the largest contribution to the total systematic error, 
which is approximately 2\% at about $\beta=2.6$. This is small, 
but larger than the statistical errors of Table~\ref{tab}. 
Fits relying on asymptotic scaling were compared with the 
conventionally used scaling fit, and the systematic error between 
the two fit forms contributes only about 0.6\% to the total 
systematic error. Our results show that one must probe rather 
deep in the scaling region to achieve systematic errors of the 
discussed variety below 1\%.

{\bf Acknowledgements:} DC was supported in part by the US Department of 
Energy (DOE) under Contract No. DE-SC0010102. Our calculations used resources
of the National Energy Research Scientific Computing Center, supported by 
the DOE under Contract No. DE-AC02-05CH11231.

\bibliography{lattice2017}

\begin{thebibliography}{17}

\bibitem{luscher_properties_2010}
M.~L\"uscher, JHEP \textbf{2010}, 1 (2010)

\bibitem{berg_dislocations_1981}
B.A. Berg, Phys. Lett. \textbf{104B}, 475 (1981)

\bibitem{bonati_comparison_2014}
C.~Bonati, M.~D'Elia, Phys. Rev. D \textbf{89}, 105005 (2014)

\bibitem{allton_lattice_1997}
C.R. Allton, Nucl. Phys. B (Proc. Suppl.) \textbf{53}, 867 (1997)

\bibitem{berg_clarke_2017}
B.A. Berg, D.A. Clarke, Phys. Rev. D \textbf{95} (2017)

\bibitem{fabricius_heat_1984}
K.~Fabricius, O.~Haan, Phys. Lett. B \textbf{143}, 459 (1984)

\bibitem{kennedy_improved_1985}
A.D. Kennedy, B.J. Pendleton, Phys. Lett. \textbf{156B}, 393 (1985)

\bibitem{adler_overrelaxation_1988}
S.L. Adler, Phys. Rev. D \textbf{37}, 458 (1988)

\bibitem{barkai_can_1982}
D.~Barkai, K.J.M. Moriarty, Comput. Phys. Commun. \textbf{27}, 105 (1982)

\bibitem{vicari_panagopoulos_2009}
E.~Vicari, H.~Panagopoulos, Phys. Rep. \textbf{470}, 93 (2009)

\bibitem{berg_markov_2004}
B.~Berg, \emph{Markov {Chain} {Monte} {Carlo} {Simulations} and {Their}
  {Statistical} {Analysis}} (World Scientific, Singapore, 2004)

\bibitem{gross_asymptotically_1973}
D.J. Gross, F.~Wilczek, Phys. Rev. D \textbf{8}, 3633 (1973)

\bibitem{politzer_reliable_1973}
H.D. Politzer, Phys. Rev. Lett. \textbf{30}, 1346 (1973)

\bibitem{caswell_asymptotic_1974}
W.E. Caswell, Phys. Rev. Lett. \textbf{33}, 244 (1974)

\bibitem{jones_two-loop_1974}
D.R.T. Jones, Nucl. Phys. B \textbf{75}, 531 (1974)

\bibitem{alle_three-loop_1997}
B.~All\'es, A.~Feo, H.~Panagopoulos, Nucl. Phys. B \textbf{491}, 498 (1997)

\bibitem{berg_asymptotic_2015}
B.A. Berg, Phys. Rev. D \textbf{92} (2015)

\end{thebibliography}
\end{document}